# Gauge Field Optics with Anisotropic Media

Fu Liu[1,2] and Jensen Li[2]*

[1]*Department of Physics and Materials Science, City University of Hong Kong, Tat Chee Avenue, Kowloon Tong, Kowloon, Hong Kong*

[2]*School of Physics and Astronomy, University of Birmingham, Birmingham B15 2TT, United Kingdom*

**By considering gauge transformations on the macroscopic Maxwell's equations, a two dimensional gauge field, with its pseudo magnetic field in the real space, is identified as tilted anisotropy in the constitutive parameters. We show that optical spin Hall effect and one-way edge states become possible simply by using anisotropic media with broadband response. The proposed gauge field also allows us to design an optical isolator based on the Aharonov-Bohm effect. Our approach will be useful in spoof magneto-optics with arbitrary magnetic fields mimicked by metamaterials with subwavelength unit cells. It also serves as a generic way to design polarization-dependent devices.**

* j.li@bham.ac.uk

Artificial magnetism from non-magnetic constituents has been showing great promise in light manipulation, particularly in the optical regime where magnetic effect is weak. As an example, metamaterials are able to create strong magnetic response to obtain optical negative refractive indices [1,2], magnetic emitters [3], perfect absorbers [4], etc. Metamaterials also enable a versatile approach to design optical devices with gradient material profiles using transformation optics (TO) approach. Prominent examples include invisibility cloaks [5-15] and optical illusion devices [16,17]. The coordinate transformation in TO draws equivalence between two material profiles and effectively changes the size and shape of an object. Another example of artificial magnetism stems from a different perspective by considering the "Lorentz force" of a photon in the geometrical optics limit. While the spin-orbit coupling of light can induce a pseudo magnetic field in the momentum space in giving rise to the optical spin Hall effect [18-20], recent research works show it is possible to realize a pseudo magnetic field in the real space instead. It complements the refractive index gradient, the pseudo electric field in the "Lorentz force", and provides an additional degree of freedom for designing optical devices through the non-conservative nature of the magnetic force. Most of the current approaches are based on a photonic lattice of coupled resonators or waveguides. Then an effective gauge field is generated by either making the coupling direction-dependent through optical path difference [21-23], dynamic modulation [24-28], or making the coupling inhomogeneous through a strain [29]. While these "bottom-up" approaches have experimentally confirmed the existence of the pseudo magnetic field in the real space, one is curious to ask whether there is a material realization of such pseudo fields, in particular we have flexible metamaterials to realize any required prescriptions of material parameters. Such an approach allows us to scale down the required structures to the subwavelength regime, and to have broadband response. The material abstraction also enables a macroscopic approach to design optical devices, similar to TO.

In this work, we consider gauge transformations on the macroscopic Maxwell's equations. While it is common to associate the refractive index to the scalar potential for photon, such a "top-down" approach allows us to directly associate also the gauge field, or the pseudo vector potential, to the constitutive parameters. Then the gradient medium required for the gauge field can be implemented by flexible metamaterials, which can have subwavelength unit cells. This abstraction, in a similar spirit to TO, enables a versatile approach in designing optical devices with pseudo magnetic fields. For simplicity, we consider 2D in-plane wave propagations on the *x-y* plane with both material parameters and fields invariant in the *z*-direction. The fields have two polarizations and can be described by a $2 \times 1$ column vector $(E_z, iH_z)$ indicating the *z*-components of the electric and magnetic fields. We begin our discussion by considering a specific class of field transformations (FT) in "rotating" polarization [30]:

$$\begin{pmatrix} E_z \\ iH_z \end{pmatrix} = \begin{pmatrix} \cos k_0\phi & -\sin k_0\phi \\ \sin k_0\phi & \cos k_0\phi \end{pmatrix} \begin{pmatrix} E_z^{(0)} \\ iH_z^{(0)} \end{pmatrix}, \tag{1}$$

where the fields with/without superscript "(0)" indicates the fields before and after transformation, $k_0 = 2\pi/\lambda_0$ is the wave number in vacuum and function $\phi$ indicates the degree of polarization rotation at each location. While applications in TO/FT usually explores the difference between the two material profiles before and after transformation (e.g. in changing size, polarization signature of an object in perception), here we purposely dismiss the polarization as an internal degree of freedom. The FT in Eq. (1) is regarded as a (Unitary) symmetry operation to keep $|E_z|^2 + |H_z|^2$ and also the Poynting vector invariant (similar to keeping $|\psi|^2$ invariant by multiplying $\psi$ with $\exp(i\phi)$ for Schrodinger's equation). From Ref. [30], such a FT induces a transformation of the material parameters and one can prove that a TO medium with equal permittivity and permeability tensor ( $\bar{\bar{\epsilon}} = \bar{\bar{\mu}} = \begin{pmatrix} \epsilon_{xx} & \epsilon_{xy} & 0 \\ \epsilon_{xy} & \epsilon_{yy} & 0 \\ 0 & 0 & \epsilon_{zz} \end{pmatrix}$ ) transforms into itself for a constant $\phi$, i.e. with global symmetry. Now, we promote the FT to be a local symmetry with spatially varying $\phi$, the TO medium is then transformed into a form of

$$\bar{\bar{\epsilon}} = \begin{pmatrix} \epsilon_{xx} & \epsilon_{xy} & A_y \\ \epsilon_{xy} & \epsilon_{yy} & -A_x \\ A_y & -A_x & \epsilon_{zz} \end{pmatrix},$$

$$\bar{\bar{\mu}} = \begin{pmatrix} \epsilon_{xx} & \epsilon_{xy} & -A_y \\ \epsilon_{xy} & \epsilon_{yy} & A_x \\ -A_y & A_x & \epsilon_{zz} \end{pmatrix}. \tag{2}$$

Finally, a further FT on such medium goes back to the same form with the gauge transformation identified as

$$A_x \to A_x + \partial_x\phi, \qquad A_y = A_y + \partial_y\phi, \tag{3}$$

where $\boldsymbol{A} = A_x\hat{x} + A_y\hat{y}$ is called the gauge field. From now on, we set $\epsilon_{xx} = \epsilon_{yy} = \epsilon_{zz} = n$, $\epsilon_{xy} = 0$ and simply call $n$ as the index of the medium. We also assume both $n$ and $\boldsymbol{A}$ are real numbers. The gauge field $\boldsymbol{A}$ can then be realized by reciprocal anisotropic metamaterials with one principal axis tilted away from the $z$-axis. We call such media as tilted anisotropic media. By comparing to the Schrodinger's equation, $\boldsymbol{A}/n^2$ is recognized as the vector/scalar potential when we decouple the Maxwell's equations (in Heaviside-Lorentz units) into pseudo spin-up/down $\psi_\pm = E_z \pm H_z$ [31]:

$$(\nabla \pm ik_0\boldsymbol{A}) \cdot (\nabla \pm ik_0\boldsymbol{A})\psi_\pm + k_0^2 n^2 \psi_\pm = 0. \tag{4}$$

Such an interpretation allows us to define a pseudo magnetic field by $\boldsymbol{B}_{\text{eff}} = \nabla \times \boldsymbol{A}$ for the photon, pointing along the z-direction. We are interested here in those media in which no gauge transformations can eliminate the gauge fields (in contrary to media considered in Ref. [30]). We note that the tilted anisotropy is not the only way to realize pseudo magnetic field in the real space (gyrotropic material is another possibility, see Supp. Info for a more complete consideration in generating a gauge transformation).

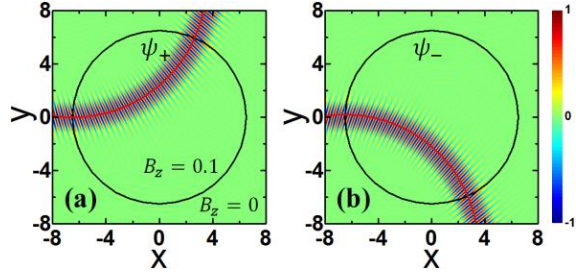

FIG. 1 (color online) The two modes bend oppositely in the pseudo magnetic field region $r \leq 6.5$ (the black circle) where $B_z = 0.1$. The mode $\psi_+/\psi_-$ bends in the (a) anti-clockwise / (b)clockwise direction.

On the other hand, if we consider the geometrical optics limit, Eq. (4) can be approximated to the ray equation (see Supp. Info for the development)

$$\frac{d}{ds}\left(n\frac{d\boldsymbol{r}}{ds}\right) = \nabla n \mp \hat{s} \times (\nabla \times \boldsymbol{A}), \qquad (5)$$

where $s$ measures the arc length of the ray, $\boldsymbol{r}$ and $\hat{s}$ denote the position vector and the propagating direction of a point on the ray. The equation is an extended version of the ray equation describing a photon propagating in a gradient index medium of isotropic indices [32]. The "Lorentz force" of the photon consists of the conventional term $\nabla n$ (from the scalar potential $n$) as a pseudo electric force. The additional term $\mp \hat{s} \times \boldsymbol{B}_{\text{eff}}$ acts as the pseudo magnetic force. A non-zero $\boldsymbol{B}_{\text{eff}}$ on the photon creates a force perpendicular to the ray direction. Such a force, having the same magnitude but opposite signs to the two pseudo spins, bends the photon trajectory.

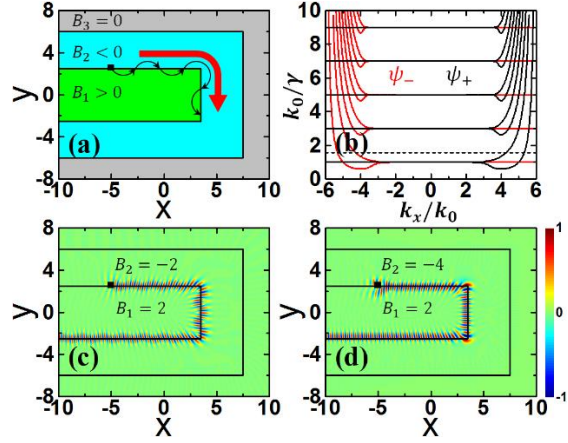

FIG. 2 (color online) (a) Edge state with ray picture (black arrows). (b) Dispersion diagram of the edge state for the two spins $\psi_+$ (black lines) and $\psi_-$ (red lines) when $\gamma = |B_1| = |B_2|$. Simulated one-way transport of the edge state ($\psi_+$) for $B_1 = 2, B_2 = -2$ (c) and $B_1 = 2, B_2 = -4$ (d).

First, we show that the materialization of the pseudo magnetic field using anisotropic media allows the demonstration of optical spin Hall effect (c.f. similar effect induced by a pseudo magnetic field in the momentum space [18-20]) when a photon travels in a region of constant $\boldsymbol{B}_{\text{eff}} = B_0 \hat{z}$ (with $B_0 = 0.1$) for $r = \sqrt{x^2 + y^2} \leq r_0$ (with $r_0 = 6.5$, the black circle in Fig. 1) while there is no pseudo magnetic field outside the region. The whole medium has a constant index $n = 1$ and we fix $\boldsymbol{A}$ (from the prescribed $\boldsymbol{B}_{\text{eff}}$) by choosing the Coulomb gauge with

$$\boldsymbol{A} = \begin{cases} 1/2 B_0 r \hat{\phi} & (r \leq r_0), \\ 1/2 B_0 r_0^2 / r \hat{\phi} & (r > r_0), \end{cases} \tag{6}$$

where $\hat{\phi}$ is the unit vector in the angular direction. We have performed full-wave simulations (using COMSOL Multiphysics) for the propagation of a Gaussian beam (of wavelength 0.4 and beam-width 1.6) within such a medium. Within the region of non-zero $\boldsymbol{B}_{\text{eff}}$, the beam with spin-up (Fig. 1 (a))/ spin down (Fig. 1(b)) is bent in the anti-clockwise /clockwise direction. It undergoes a circular motion with radius of $n/B_{\text{eff}}$ (=10, the red curve). Within an area of zero magnetic field (when the beam exits), the modes exhibit no bending forces and will keep propagation with negligible deflection. Such bending effect is broadband in nature (see Supp. Info.).

The magnetic force (apart from the "electric" force from index gradient) allows wave guiding with a very different mechanism. For example, with an interface between regions of pseudo magnetic fields of opposite signs, the photon can be guided on the interface, as an edge state [25], with the two opposite bending forces illustrated in the ray picture (see Fig. 2(a)). It supports a one-way transport mode, which can travel around sharp corners without backscattering. We have simulated such a phenomenon in Fig. 2(c). The whole domain (with constant $n = 1$) has three different regions with constant pseudo magnetic fields of $B_1 = 2$ (the inner region), $B_2 = -2$ (the ring) and $B_3 = 0$ (outside). The required materialization of the vector potential is again fixed (numerically) by the Coulomb gauge [0]. A point source located at a position near the interface between opposite magnetic fields (the solid dot) emits spin-up photon (with a free space wavelength of 2), which travels to the right and bends around the corners without reflection. We have also obtained the dispersion diagram, shown in Fig. 2(b), of this edge state on the flat surface (see Supp. Info for more details) where $\gamma = |B_1| = |B_2|$ is the magnitude of the pseudo magnetic field on the two sides. Black/red curves show the spin-up/spin-down modes. They are mirror copies ($k_x \to -k_x$) of each other, resulting from the time-reversal symmetry respected by our system. The flat bands (largely negative/positive $k_x$ for spin-up/spin-down) with $k_0/\gamma$ around odd numbers are the odd/even mode combinations of the cyclotron modes on the two sides approaching the so called Landau levels with little (evanescent wave) coupling. They evolve and with more prominent splitting (e.g. at larger values of $k_x$ for the spin-up) to form the waveguide modes. The upper/lower mode has odd/even symmetry about the interface. The horizontal dashed line in Fig. 2(b) indicates the frequency for the simulation in Fig. 2(c) with free-space wavelength $\lambda_0 = 2$. For spin-up excitation, it has two modes with $k_x \approx 4.3k_0$ and $k_x \approx 5.3k_0$. They beat together, forming the wiggling pattern, and propagate only to the right as shown in Fig. 2(c). We note that the one-way transport is broadband in frequency response. As the excitation frequency changes, we can still find edge modes in the dispersion diagram for $k_0/\gamma > 0.59$. Furthermore, the one-way transport is also possible when $|B_1| \neq |B_2|$. Fig. 2(d) shows the asymmetric case when $B_2$ is changed to $-4$. The two sets of flat bands (the Landau levels) on the two sides split (see Supp. Info), the working frequency (the horizontal dashed line) in this case only cuts the band of lower guiding mode of even symmetry, so that the one-way transport is observed without beating (in Fig. 2(d)). We note that, the phenomenon for the spin-down mode is completely opposite (one way transport to the left) for the same medium.

Apart from modifying the photon trajectory, there is an additional geometric phase $\phi_g$ along propagation in the case of a non-zero pseudo magnetic field. For a round trip, $\phi_g$ is given by

$$\phi_g = \mp \oint_{\partial \Sigma} k_0 \boldsymbol{A} \cdot d\boldsymbol{r} = \mp k_0 \iint_{\Sigma} \boldsymbol{B}_{eff} \cdot d\boldsymbol{\Sigma}, \tag{7}$$

where $\partial\Sigma$ denotes the ray trajectory and $\Sigma$ denotes the enclosed area (with second equality obtained by Stoke's theorem). The upper (lower) sign corresponds to the mode $\psi_+$ ($\psi_-$). In other words, the propagation phase now becomes path-dependent, as the so called photonic Aharonov-Bohm effect in this tilted anisotropic media (another approach is based on dynamic modulation, see Ref. 24).

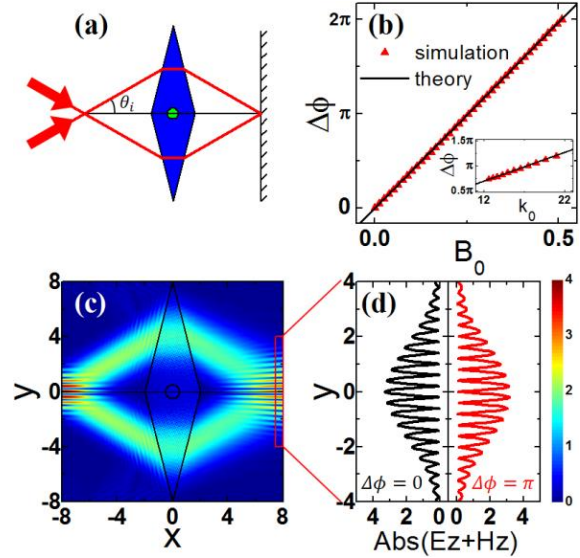

FIG. 3 (color online) Aharonov-Bohm effect with tilted anisotropic media. (a) Two spin-up beams focused by a prism to the screen on the right for interference. The green circle is filled with constant pseudo magnetic field $B_0\hat{z}$. (b) Simulation results (red triangles) comparing with analytic formula (black lines) for the phase difference ($\Delta\phi$) between the two incident beams at focal point on screen. (c) The interference pattern $abs(E_z + H_z)$ of the two beams when $B_0 = 0$. (d) The interference $abs(E_z + H_z)$ on the screen when $B_0 = 0$ (giving $\Delta\phi = 0$, black curve, peak at $y = 0$) and when $B_0 = 0.255$ ($\Delta\phi = \pi$, red curve, dip at $y = 0$).

In the following, we will first show such a path-dependent phase through interference effect and then employ it to design an optical isolator for a particular spin mode. Fig. 3(a) shows our schematic setup to interfere two photon paths with the same starting and end points in a background medium of index $n = 1$. Two spin-up beams (red arrows) with incidence angles $\theta_i = \pm 30°$ and equal initial phase are focused by a prism (blue color, with index $n = 2.866$) to the screen on the right where an interference pattern is recorded. The pseudo magnetic field of the whole domain is zero except the small region (the green circular disk in the center of the prism with radius $r_0 = 0.5$) where a constant pseudo magnetic field of $B_0\hat{z}$ is applied. Fig. 3(c) shows the full-wave simulation of the interference between the two beams (with wavelength 0.4 and beam width 2) when $B_0 = 0$. The two beams meet at $x = 8$ with the interference pattern shown in Fig. 3(d). There is no phase shift between the two beams with a constructive interference peak at $y = 0$, indicated as the black curve ($\Delta\phi = 0$). Now, the pseudo magnetic field is turned on ($B_0 = 0.255$, with the materialization of vector potential fixed by Eq. (6)). While the beam trajectory stays nearly the same without exhibiting a magnetic force, the interference pattern now is shifted with destructive interference at $y = 0$, i.e. $\Delta\phi = \pi$, being shown as the red curve in Fig. 3(d). From the shift of interference pattern $\Delta y_{\text{peak}}$, we can derive $\Delta\phi$ (using $\Delta y_{\text{peak}} = \frac{\lambda_0}{4\pi \sin\theta_i}\Delta\phi$), the phase difference between the two paths, which is plotted in Fig. 3(b) using red triangular symbols. They are found in good agreement with the analytic prediction from Eq. (7), $\Delta\phi = k_0 B_0 \pi r_0^2$ when we vary the magnitude of $B_0$ or vary the wavenumber $k_0$. The significance of Eq. (7) is that the geometric phase is only related to the enclosed magnetic flux even though there is no pseudo magnetic field along the ray trajectory. In another perspective, the phase difference is created due to the additional anisotropy, the vector potential $\mathbf{A}$, we put into the medium. If this vector potential has a non-conservative nature, in contrary to all previously demonstrated TO and FT media, we can have a path-dependent phase, which is not obvious for the usual geometrical optics in anisotropic media. Based on this, we can actually design devices with non-trivial profile of $\mathbf{A}$ (instead of generating $\mathbf{A}$ from specified pseudo magnetic fields) directly.

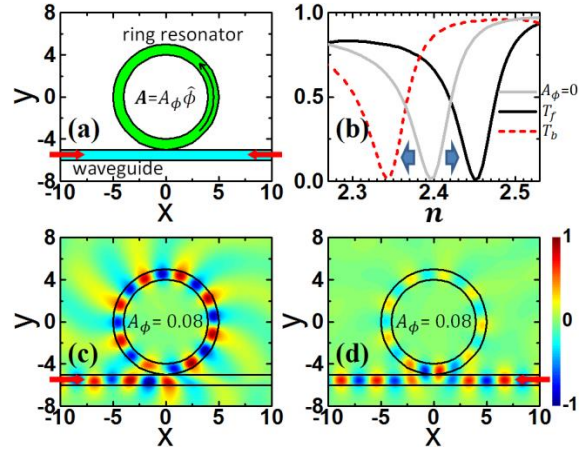

FIG. 4 (color online) Optical isolator for spin-up mode ($\psi_+$). (a) Schematic of design: a ring resonator with vector potential $\mathbf{A} = A_\phi \hat{\phi}$ coupled to a background waveguide with incidence either from left or right. (b) Power transmittance for both forward ($T_f$, black solid line) and backward ($T_b$, red dashed line) incidence versus the waveguide index $n$ when $A_\phi = 0.08$. The gray line shows the case with symmetric transmittance when $A_\phi = 0$. (c, d) Field patterns ($\psi_+ = E_z + H_z$) for forward incidence (c) and backward incidence (d) when $A_\phi = 0.08$ and $n = 2.45$.

Finally, we employ the path-dependent geometric phase to design an optical isolator for the spin-up mode by coupling a ring resonator with a vector potential $\boldsymbol{A}$ to a straight waveguide (Fig. 4(a)). The ring resonator has an inner/outer radius of 4/5, touching a straight waveguide of unit height. The free-space wavelength here is fixed at 5.5 while the index of the waveguide and the ring is designed to resonate at $n = 2.4$ so that a guiding mode initially travelling along the straight waveguide is completely reflected and scattered (same for both forward incidence and backward incidence), with a transmission dip (gray color) when we vary $n$ in Fig. 4(b). Then, we add a constant $A_\phi = 0.08$ (as tilted anisotropy) pointing in the anticlockwise direction wherever inside the ring. It adds/subtracts a geometric phase for the spin-up modes propagating in the clockwise/anti-clockwise direction. This splits the resonating condition of the ring resonator, or equivalently the forward (left to right) and backward transmission curve, as shown as the black and red curves in Fig. 4(b). Then, when we shift to work at $n = 2.45$ (the position of the forward transmission dip), the backward and forward spin-up guiding modes have nearly unit and zero transmission ($T_b \cong 0.95 / T_f \cong 0.01$) with a large contrast. We note that the system here is discussed within wave propagation of spin-up (with decoupled spin-up/spin-down mode propagation in this system). We have assumed the device is working with spin-up wave so that the isolator functionality is defined with respect to the fundamental spin-up mode of the waveguide in both forward and backward directions. It is in contrary to the usual discussion of isolators that the backward propagation mode is the time-reversed copy of the forward propagation mode. Our system is still reciprocal and the time reversed copy of the forward spin-up propagation mode actually goes to the backward spin-down propagation mode.

In conclusion, we have proposed a scheme to realize the gauge field and the pseudo magnetic field for photon propagation using reciprocal anisotropic media. Such a materialization using anisotropic media allows us to design optical devices enabled by the additional bending power from pseudo magnetic fields or the path-dependent geometric phase introduced by a non-trivial gauge field. As illustrations, we have demonstrated one-way transport of edge states and an optical isolator using tilted anisotropic media.

**References**

1. V. M. Shalaev et.al., *Opt. Lett*. 30, 3356 (2005).
2. S. Zhang et.al., *Phys. Rev. Lett*. 95, 137404 (2005).
3. M. L. Povinelli, S. G. Johnson, J. D. Joannopoulos and J. B. Pendry, *Appl. Phys. Lett*. 82, 1069 (2003).
4. N. I. Landy et.al., *Phys. Rev. Lett*. 100, 207402 (2008).
5. J. B. Pendry, D. Schurig, and D. R. Smith, *Science* 312, 1780 (2006).
6. U. Leonhardt, *Science* 312, 1777 (2006).
7. D. Schurig, J. Mock, B. Justice, S. A. Cummer, J. Pendry, A. Starr, and D. Smith, Science 314, 977 (2006).
8. R. Liu, C. Ji, J. Mock, J. Chin, T. Cui, and D. Smith, Science 323, 366 (2009).
9. J. Valentine, J. Li, T. Zentgraf, G. Bartal, and X. Zhang, Nature Mater. 8, 568 (2009).
10. L. H. Gabrielli, J. Cardenas, C. B. Poitras, and M. Lipson, Nature Photon. 3, 461 (2009).
11. T. Ergin, N. Stenger, P. Brenner, J. B. Pendry, and M. Wegener, Science 328, 337 (2010).
12. X. Chen, Y. Luo, J. Zhang, K. Jiang, J. B. Pendry, and S. Zhang, Nature Commun. 2, 176 (2011).
13. B. Zhang, Y. Luo, X. Liu, and G. Barbastathis, Phys. Rev. Lett. 106, 33901 (2011).
14. J. Zhang, L. Liu, Y. Luo, S. Zhang, and N. A. Mortensen, Opt. Express 19, 8625 (2011).
15. M. Gharghi, C. Gladden, T. Zentgraf, Y. Liu, X. Yin, J. Valentine, and X. Zhang, Nano Lett. 11, 2825 (2011).
16. Y. Lai, J. Ng, H. Y. Chen, D. Z. Han, J. J. Xiao, Z. Q. Zhang, and C. T. Chan, *Phys. Rev. Lett.* **102**, 253902 (2009).
17. C. Li, X. Meng, X. Liu, F. Li, G. Fang, H. Chen, and C. T. Chan, *Phys. Rev. Lett.* **105**, 233906 (2010).
18. M. Onoda, S. Murakami and N. Nagaosa, Phys. Rev. Lett. 93, 083901 (2004).
19. O. Hosten and P. Kwiat, Science 319, 787 (2008).
20. K. Y. Bliokh, *J. Opt. A:Pure Appl. Opt.* **11**, 094009 (2009).
21. M. Hafezi et.al., Nature Phys. 7, 907 (2011).
22. R O. Umucalilar and I. Carusotto, Phys. Rev. A 84, 043804 (2011).
23. M. Hafezi, S. Mittal, J. Fan, A. Migdall and J. M. Taylor, Nature Photon. 7, 1001 (2013).
24. K. Fang, Z. Yu, and S. Fan, *Phys. Rev. Lett.* **108**, 153901 (2012).
25. K. Fang, Z. Yu, and S. Fan, *Nat. Photonics* **6**, 782 (2012).
26. M. C. Rechtsman et.al., Nature 496, 196 (2013).
27. L. D. Tzuang, K. Fang, P. Nussenzveig, S. Fan and M. Lipson, Nature Photon. 8, 701 (2014).
28. Q. Lin, S. Fan, *Phys. Rev. X* **4**, 031031 (2014).
29. M. C. Rechtsman et.al., *Nature Photon*. 7, 153 (2013).
30. F. Liu, Z. Liang, and Jensen Li, *Phys. Rev. Lett.* **111**, 033901 (2013).
31. A. B. Khanikaev et.al., Nature Mater. 12, 233 (2013).
32. M. Born and E. Wolf, *Principles of optics*, (Cambridge University Press, Cambridge, 2005).

Here, we solve the profile of vector potential $\boldsymbol{A}$ by numerically solving a Laplace equation obtained by substituting $\nabla \times \boldsymbol{A} = B_z \hat{z}$ with $\boldsymbol{A} = \partial_y \Phi \hat{x} - \partial_x \Phi \hat{y}$ and by setting zero $\Phi$ at the outside boundary of simulation domain to approximately ensure $\boldsymbol{A} = \boldsymbol{0}$ at infinity, see Supp. Info for more details.